\documentclass[prd, twocolumn, superscriptaddress, nofootinbib, amsmath, amssymb,
    aps, floatfix, 10pt
]{revtex4-2}
\usepackage[utf8]{inputenc}
\usepackage{graphicx,xcolor,setspace}
\usepackage[
    colorlinks=true,
    linkcolor=blue,
    urlcolor=blue,
    citecolor=blue
]{hyperref}

\usepackage[capitalise]{cleveref}
\usepackage{siunitx}
\usepackage{orcidlink}

\IfFileExists{lmodern.sty}{\usepackage{lmodern}}{}
\usepackage{babel}
\usepackage{hyperref}

\usepackage[pagewise]{lineno}
\newcommand{\lya}{Lyman-$\alpha$ } 
 
\newcommand{\lcdm}{$\Lambda$CDM}
\newcommand{\DP}{$A'$ DM}
\newcommand{\mnras}{Mon. Not. Roy. Astron. Soc.}
\newcommand{\apjl}{Astrophys. J. Lett.}
\newcommand{\pasp}{Publ. Astron. Soc. Pacific.}
\newcommand{\jcap}{JCAP}
\newcommand{\pasa}{PASA}

\DeclareSIUnit\eVc{\eV\per\clight\tothe{2}}
\DeclareSIUnit\clight{\text{\ensuremath{c}}}

\makeatother

\begin{document}

\begin{flushright}
    CERN-TH-2024-166
\end{flushright}

\title{%The Dark Sherwood Forest \\
Constraints on dark photon dark matter from Lyman-$\alpha$ forest simulations and an ultra-high signal-to-noise quasar spectrum 
}

%%%%%%%%%%%%
% AUTHORS
%%%%%%%%%%%%

\author{Andrea Trost\,\orcidlink{0000-0002-5959-5964}}
\email{andrea.trost@inaf.it}
\affiliation{Dipartimento di Fisica dell’Università di Trieste, Sezione di Astronomia, Via G.B. Tiepolo, 11, I-34143 Trieste, Italy.}
\affiliation{INAF - Osservatorio Astronomico di Trieste, Via G. B. Tiepolo 11, I-34143 Trieste, Italy}
\affiliation{INFN -  Sezione di Trieste, Via Valerio 2, I-34127 Trieste, Italy}

% \affiliation{INAF – Osservatorio Astronomico di Trieste, via G.B. Tiepolo, 11, I-34143 Trieste, Italy}
% \affiliation{INFN – National Institute for Nuclear Physics, via Valerio 2, I-34127 Trieste, Italy}
\author{James S. Bolton\,\orcidlink{0000-0003-2764-8248}}
\email{james.bolton@nottingham.ac.uk}
\affiliation{School of Physics and Astronomy, University of Nottingham, University Park, Nottingham, NG7 2RD, UK}

\author{Andrea Caputo\,\orcidlink{0000-0003-1122-6606}} \email{andrea.caputo@cern.ch}
\affiliation{Theoretical Physics Department, CERN, 1211 Geneva 23, Switzerland}
\affiliation{Dipartimento di Fisica, ``Sapienza'' Università di Roma \& Sezione INFN Roma1, Piazzale Aldo Moro
5, 00185, Roma, Italy}

\author{Hongwan Liu\,\orcidlink{0000-0003-2486-0681}}
% \email{hongwan@bu.edu}
\affiliation{Physics Department, Boston University, Boston, MA 02215, USA}
\affiliation{Kavli Institute for Cosmological Physics, University of Chicago, Chicago, IL 60637}
\affiliation{Theoretical Physics Department, Fermi National Accelerator Laboratory, Batavia, IL 60510}

\author{Stefano Cristiani\,\orcidlink{0000-0002-2115-5234}}
% \email{stefano.cristiani@inaf.it}
\affiliation{INAF - Osservatorio Astronomico di Trieste, Via G. B. Tiepolo 11, I-34143 Trieste, Italy}
\affiliation{INFN - Sezione di Trieste, Via Valerio 2, I-34127 Trieste, Italy}
\affiliation{IFPU - Institute for Fundamental Physics of the Universe, Via Beirut 2, I-34151 Trieste, Italy}

\author{Matteo Viel\orcidlink{0000-0002-2642-5707}}
% \email{viel@sissa.it}
\affiliation{SISSA - International School for Advanced Studies, Via Bonomea 265, I-34136 Trieste, Italy}
\affiliation{IFPU - Institute for Fundamental Physics of the Universe, Via Beirut 2, I-34151 Trieste, Italy}
\affiliation{INFN - Sezione di Trieste, Via Valerio 2, I-34127 Trieste, Italy}
\affiliation{INAF - Osservatorio Astronomico di Trieste, Via G. B. Tiepolo 11, I-34143 Trieste, Italy}

\begin{abstract}
%\blue{$<$ 200 words}
The ultralight dark photon is a well-motivated, hypothetical dark matter candidate. In a dilute plasma, they can resonantly convert into photons, and heat up the intergalactic medium between galaxies. In this work, we explore the dark photon dark matter parameter space by comparing synthetic Lyman-$\alpha$ forest data from cosmological hydrodynamical simulations to observational data from VLT/UVES of the quasar HE0940-1050 ($z_{\rm em}=3.09$). We use a novel flux normalization technique that targets under-dense gas, reshaping the flux probability distribution. Not only do we place robust constraints on the kinetic mixing parameter of dark photon dark matter, but notably our findings suggest that this model can still reconcile simulated and observed Doppler parameter distributions of $z\sim0$ Lyman-$\alpha$ lines, as seen by HST/COS. This work opens new pathways for the use of the \lya forest to explore new physics, and can be extended to other scenarios such as primordial black hole evaporation, dark matter decay, and annihilation.

\end{abstract}
\maketitle

\section{Introduction}
% \blue{Journal guidelines -- Nature Communications:}\\
% \blue{Intro + results + discussion $<$ 5000 words }\\
% \blue{Methods < 3000 words}
% \blue{Fig + Tables $\le$ 10}\\
% \blue{$\le$ 70 references}\\
% \blue{Fig legends $<$ 350 words}\\

Dark matter (DM) comprises roughly 27$\%$ of the energy content of the Universe, while regular matter, which we see all around us, makes up only 5$\%$~\cite{Planck:2018vyg}. What is dark matter? How was it formed in the early Universe? And how does it interact with regular matter? Despite decades of theoretical and experimental efforts, these pressing questions remain unanswered. To date, the most widely studied form of dark matter has been the Weakly Interacting Massive Particle (WIMP)~\cite{Steigman:1984ac, Primack:1988zm}, theoretically motivated by the hierarchy problem, which suggests new physics around the weak scale. More recently, however, the community has made a broader effort to explore scenarios of DM significantly lighter than the weak scale. 
Two well-motivated, ultralight DM candidates are the axion~\cite{Peccei:1977hh, Weinberg:1977ma, Wilczek:1977pj} and the dark photon ($A'$)~\cite{Holdom:1985ag}. In particular, $A'$ are hypothetical new massive vectors that are a minimal extension of the Standard Model (SM) of particle physics. This new particle couples to the ordinary photon via the kinetic mixing operator, one of the few marginal operators that we can write down to couple the SM to a dark sector.

The coupling between ordinary and dark photons gives rise to a rich phenomenology, and recent years have seen the field bloom with new detection ideas. A vast experimental program has been developed to \textit{directly detect} dark photon DM in the laboratory~\cite{Caputo:2021eaa}, using cavities, dielectric disks, dish antennae, plasmas, LC circuits, electric-field radios, small electric-dipole antennas. However, laboratory experiments in the quest for dark matter are typically limited by their size $L_\text{exp}$, and have difficulty accessing $A'$ masses smaller than $m_{A'} \lesssim \hbar/(L_\text{exp}c^3)$. Astrophysical and cosmological probes are generally better suited for smaller masses, since they provide longer baselines for the interaction between \DP~and photons. Some observables of this kind include spectral distortions to the cosmic microwave background (CMB)~\cite{Mirizzi:2009iz, McDermott:2019lch, Witte:2020rvb, Caputo:2020bdy, Caputo:2020rnx}, the solar corona~\cite{An:2020jmf, An:2023wij}, and the Earth's ionosphere~\cite{Beadle:2024jlr}, among others. In this work we contribute to this effort and thoroughly investigate the impact of \DP~on the cosmic web.

In the current paradigm, after the reionization epoch~\cite{McQuinn2016_review}, the diffuse baryonic gas scattered in the Universe -- the intergalactic medium (IGM) -- is mainly composed of ionized hydrogen (HII) with traces of neutral hydrogen (HI) and helium, whose abundances depend on redshift and the ionizing background. The IGM is a powerful tracer of the underlying DM distribution on scales above the gas pressure smoothing scale of $\sim 100$ proper kpc. 
The main observable feature of this diffuse gas is the series of absorption lines it imprints on the spectra of luminous, distant quasars. Light from these sources travelling through the IGM is resonantly scattered along the line of sight as photons redshift into the frequency of atomic transitions in the rest frame of the intervening gas. 
The \lya transition of neutral hydrogen ($n=1\rightarrow2$, $\lambda=1215.67 \text{\AA}$) is a prime example, as it generates a crowded series of lines -- the so-called \lya forest -- spanning a redshift range $\Delta z\sim0.5 - 1$, depending on emission redshift, along each quasar sightline.

The power spectrum of the \lya forest has been extensively used in cosmology to probe the large-scale structure of the Universe~\cite{SDSS:2004kjl, Slosar2011}, cosmic reionization~\cite{Becker2015, Becker2015PASA}, and the thermal history of the IGM~\cite{Schaye1999,Boera2014}. It can effectively constrain $\Lambda$CDM cosmological parameters, including the energy density of dark energy~\cite{Viel:2002gn, SDSS:2004kqt}. 
It has also been used to study beyond-$\Lambda$CDM cosmologies that modify the structure of the cosmic web at small scales, such as warm dark matter (WDM)~\cite{Viel:2005qj, Viel:2013fqw, Irsic2017New}, sterile neutrinos~\cite{Abazajian2001, Seljak:2006qw}, 
fuzzy DM \cite{Hu:2000ke, Hui:2016ltb, Armengaud:2017nkf, Irsic:2017yje}
and primordial black hole DM~\cite{Carr2016pbh, Murgia:2019duy}.

Although almost all studies so far have focused on the use of the power spectrum, or the two point correlation function \cite{DESI2024} (i.e. the power spectrum's Fourier transform), we argue that single-point statistics are also extremely sensitive to new physics. In fact, the \lya forest is an exquisite tracer of the thermodynamic state of the gas at $z<5$, as the line's profile depends on the absorbing medium's temperature, which makes it an excellent probe of exotic energy injection during this epoch. Intriguingly, recent results by different groups~\cite{Viel2017,Gaikwad2017,Burkhart:2022ygp} have shown hints of a discrepancy in the temperature of the IGM at $z\sim0$ that cannot be reconciled with AGN feedback mechanisms and/or photoheating by ultraviolet (UV) photons. These works suggest the presence of an unaccounted-for source of heating at $z\sim2$. 

Such a source could be provided by DM interactions. If DM is made of ultralight dark photons $A'$, these can \textit{resonantly} convert into soft photons when suitable conditions are met in dilute plasma. The soft photons would then be efficiently absorbed by the plasma, heating it up~\cite{McDermott:2019lch}. If the conditions for the conversions to happen are realized at $z < 5$, this heating will strongly impact the \lya forest. Initially, crude limits were derived on the kinetic mixing parameter by simply requiring that the amount of heat injected not exceed 1 eV per baryon~\cite{Caputo:2020bdy,Caputo:2020rnx,Witte:2020rvb}. 

Recently, however, Ref.~\cite{Bolton:2022hpt} made a significant step forward by implementing the effect of \DP~heating into full-fledged, cosmological hydrodynamical simulations of the cosmic web. By comparing changes to the predicted IGM temperature with the same quantity inferred from \lya forest measurements, the authors found that \DP~could indeed explain the low-redshift \lya heating anomaly, and predicted that it could impact the \lya forest at higher redshifts. 

Motivated by this important finding, we conduct an extensive, detailed, and up-to-date study of the impact of ultralight \DP~on the IGM at different epochs using, for the first time, the full \lya flux from observational data at $z \sim 3$.  
We use new simulations with significantly higher resolution, which solve the \textit{complete} non-equilibrium photoionization equations (unlike in Ref.~\cite{Bolton:2022hpt}) and incorporate the latest UV photoheating rates \cite{Puchwein2019}. We perform 25 simulations with \DP~(the largest set to date) and scan over a wide range of parameter space. Furthermore, we adopt a novel flux normalization technique which targets under-dense gas and, for the first time, we compare directly with the transmitted flux using a forward model at $z\sim3$, rather than using the derived measurements of the IGM temperature.
In so doing, we set leading constraints on the kinetic mixing parameter of ultralight $A'$ DM that are significantly more robust than previous results.

\section{Results}

\subsection{Conversion and absorption of dark photon dark matter in the IGM}
The Lagrangian describing photons and dark photons is~\cite{Holdom:1985ag}
\begin{equation}
    \mathcal{L}_{\gamma A'} = -\frac{1}{4}F^2_{\mu\nu} - \frac{1}{4}(F'_{\mu\nu})^2-\frac{\epsilon}{2}F^{\mu\nu}F'_{\mu\nu}+\frac{1}{2}m^2_{A'}(A'_\mu)^2,
\end{equation}
where $F$ and $F'$ are the field strength tensors for the ordinary and dark photons, respectively, $\epsilon$ is the dimensionless kinetic mixing parameter, and $m_{A'}$ is the mass of the dark photon. Various models propose that this new particle constitutes the entirety of dark matter, produced via a misalignment mechanism with a nonminimal coupling to the Ricci scalar~\cite{Arias:2012az, Graham:2015rva, AlonsoAlvarez:2019cgw}, via tachyonic instabilities with misaligned axions~\cite{Agrawal:2018vin, Co:2018lka, Bastero-Gil:2018uel}, or the decay of topological defects~\cite{Long:2019lwl} (see Refs.~\cite{East:2022rsi,Cyncynates:2023zwj,Cyncynates:2024yxm} for more critical studies on production mechanisms). Throughout this paper, we will assume that DM is completely made up of ultralight dark photons. 

The kinetic mixing term induces oscillations of dark photons into photons and vice versa. The SM photon, in the presence of a charged plasma such as the ionized IGM, gains an effective plasma mass that depends mainly on the local free electron density $n_e$. In the limit of a cold and non-degenerate plasma (a very good approximation for the IGM at low redshift), the effective plasma mass at every point in space $\textbf{x}$ and redshift $z$ is given by
\begin{equation}
    m^2_\gamma \simeq 1.4\times10^{-21}~\unit{\eV\tothe{2}\per\clight\tothe{4}}
    \left(\frac{n_e(z,\textbf{x})}{\text{cm}^{-3}}\right).\,
\end{equation}
Whenever and wherever the standard photon effective plasma mass equals the $A'$ mass, i.e.\ $m^2_\gamma(z,\textbf{x}) = m^2_{A'}$, the transition between the two particle species is resonantly enhanced. Assuming that the resonant condition is met at a time $t_{\rm res}$, the transition probability reads~\cite{Caputo:2020bdy, Caputo:2020rnx}
\begin{equation}
    P_{A'\rightarrow\gamma}(\textbf{x},t_{\rm res}) = \pi \epsilon^2 \frac{m_{A'}c^2}{\hbar}\left| \frac{d\ln m_{\gamma}^2(\textbf{x},t)}{dt}\right|^{-1}_{t=t_{\rm res}}.
\end{equation}

Low mass dark photons ($m_{A'}\sim10^{-15}-10^{-12}~\unit{\eVc}$) convert to low-frequency ($\sim 1-10^3 \, \text{Hz}$)  photons which are rapidly absorbed by the IGM through free-free absorption. Ref.~\cite{Bolton:2022hpt} estimated the mean free path of the newly produced photons to be much smaller than the average \lya absorber, shown to be $\sim \SI{100}{kpc}$~\cite{Dodorico98, Cristiani24}, and therefore we can assume that the photon absorption is localized to the region where the resonant condition is met.  

Due to this absorption, the oscillation between species results in a net flow of energy into the IGM, and an increase in the gas temperature at the density at which the transition happens. The energy injection per baryon left by the transition can be approximated as~\cite{Bolton:2022hpt}
\begin{equation}\label{eq:E-dp_ph}
    E_{A'\rightarrow\gamma}\sim 2.5~\unit{\eV}
\left(\frac{\epsilon_{-14}}{0.5}\right)^2
\left(\frac{m_{-13}}{0.8}\right)
\left(\frac{1+z_{\rm res}}{3}\right)^{-3/2},
\end{equation}
where $z_\text{res}$ is the redshift of the resonance, $\epsilon_{-14}\equiv\epsilon/10^{-14}$, and $m_{-13}\equiv m_{A'}/\left(10^{-13} ~\unit{\eVc}\right)$. 
An approximate expression for $z_\text{res}$ is 
\begin{equation}\label{eq:zres}
    1 + z_\text{res} \sim 2.75 \left( \frac{m_{-13}}{0.8} \right)^{2/3} \frac{1}{\Delta_b^{1/3}} \,,
\end{equation}
where $\Delta_b = \rho_b / \bar{\rho}_b$ is the relative overdensity of gas with respect to the mean. 
Resonance conversions occur in underdense regions first before overdense ones. Throughout the remainder of this paper, we shall refer to $\Delta_b$ simply as $\Delta$ for brevity, unless otherwise specified.

\subsection{\DP~heating in cosmological hydrodynamical simulations}
% To study the impact of \DP~on the \lya forest, we adopt state-of-the-art models based on the Sherwood-Relics suite of cosmological hydrodynamical simulations \cite{Puchwein:2022wvk,Bolton2017} performed using P-Gadget-3 \cite{Springel2005}.  We modify these simulations to include energy injection into the baryonic gas particles due to the $A' \rightarrow \gamma$ transition when the gas particles satisfy the resonant condition $m^2_\gamma = m^2_{A'}$. 
% We thoroughly investigate the \DP~parameter landscape by running several models with varying \DP~mass and kinetic mixing. 

To study the impact of \DP~on the \lya forest, we adopt high-resolution cosmological hydrodynamical simulations of the \lya forest performed with a modified version of P-Gadget-3, which is an extended version of the publicly available Gadget-2 code \cite{Springel2005}.   We use the code originally modified for the Sherwood-Relics simulation project \cite{Puchwein:2022wvk,Bolton2017}, where it was demonstrated these models reproduce the observed \lya forest at $2\leq z \leq 5$.    In this work our fiducial simulation box size is $L = 40 h^{-1}\unit{Mpc}$ with $2\times 1024^3$ DM and gas particles, giving particle masses of $M_{\rm dm}=4.3\times 10^{6}h^{-1}\rm\,M_{\odot}$ and $M_{\rm gas} = 7.97\times 10^{5}h^{-1}\rm\,M_{\odot}$.  We demonstrate later that this choice yields a well-converged \lya forest tPDF. 

The simulations were all started at $z=99$, with initial conditions generated on a regular grid using a $\Lambda \rm CDM$ transfer function generated by CAMB \cite{Lewis2000}.  The simulations were evolved to $z=2$, and the gas thermo-chemistry was obtained by solving the full network of non-equilibrium photo-ionization equations for hydrogen and helium \cite{Puchwein2015}.  Ionization and heating by UV photons is followed using the empirically-calibrated synthesis UV background model of Ref.~\cite{Puchwein2019}, with photo-heating rates adjusted by a factor of $0.8$ to match more recent   IGM temperature measurements presented in Ref.~\cite{Gaikwad2021}.  This adjustment is equivalent to using a slightly softer UV background spectrum relative to the original model used by Ref.~\cite{Puchwein2019}; it lowers the IGM temperature by a few thousand degrees Kelvin.  The exact spectral shape of the UV background is uncertain, so this approach ensures good agreement with more recent observational constraints on the IGM gas temperature that were unavailable when Ref.~\cite{Puchwein2019} was published.  Note also that the redshift evolution of the UV background model remains unchanged.   
All gas particles in the simulations with density $\Delta>10^{3}$ and temperature $T<10^{5}\rm\,K$ are converted into collisionless star particles \cite{Viel2004}.  We assume an underlying $\Lambda\rm CDM$ cosmology with the following parameters: $\Omega_{\rm m}=0.308$, $\Omega_\Lambda=0.692$, $h=0.678$, $\Omega_{\rm b}=0.0482$, $\sigma_8=0.829$, $n=0.961$, and a helium mass fraction $Y_{\rm p}=0.24$. 

We modify these simulations to include energy injection into the baryonic gas particles due to $A' \rightarrow \gamma$ following Ref.~\cite{Bolton:2022hpt}. This is included \emph{in addition} to photo-heating from our baseline UV background model, by applying a direct energy injection into the gas particles (following Eq. \ref{eq:E-dp_ph}) when the resonant condition $m_\gamma^2 = m^2_{A'}$ is met.  
We performed $25$ simulations spanning the $A'$ parameter space, with masses $-13.4 \le \log(m_{\rm A'}/\unit{\eVc})\le -11.9$ and kinetic mixing $-15.0 \le \log\epsilon \le -13.5$.  Table \ref{tab:table1} reports the $A'$ parameters considered. We also performed $4$ additional simulations with no $A'$ heating to test for convergence with mass resolution and box size.

\begin{table}[t]
\begin{ruledtabular}
\begin{tabular}{ll}
\textrm{$\log(m_{\rm A'}/\unit{\eVc}) $}&
\textrm{$\log \epsilon$}\\
\colrule
$-13.40$ & $-15.00,\; -14.50,\; -14.00$ \\
$-13.15$ & $-14.75,\; -14.15,\; -14.00, \; -13.50$ \\ 
$-12.90$ & $-15.00,\; -14.50,\; -14.15,\; -14.00$ \\
$-12.65$ & $-14.75,\; -14.15,\; -14.00$ \\
$-12.40$ & $-15.00,\; -14.50,\; -14.15,\; -14.00$ \\
$-12.15$ & $-14.75,\; -14.15,\; -14.00,\; -13.50$ \\ 
$-11.90$ & $-15.00,\; -14.50,\; -14.00$

\end{tabular}
\end{ruledtabular}
\caption{\label{tab:table1}%
Summary of the \DP~parameters assumed in the hydrodynamical simulations used in this work.
}
\end{table}

\begin{figure*}[t!]
    \centering
    \includegraphics[width=.9\textwidth]{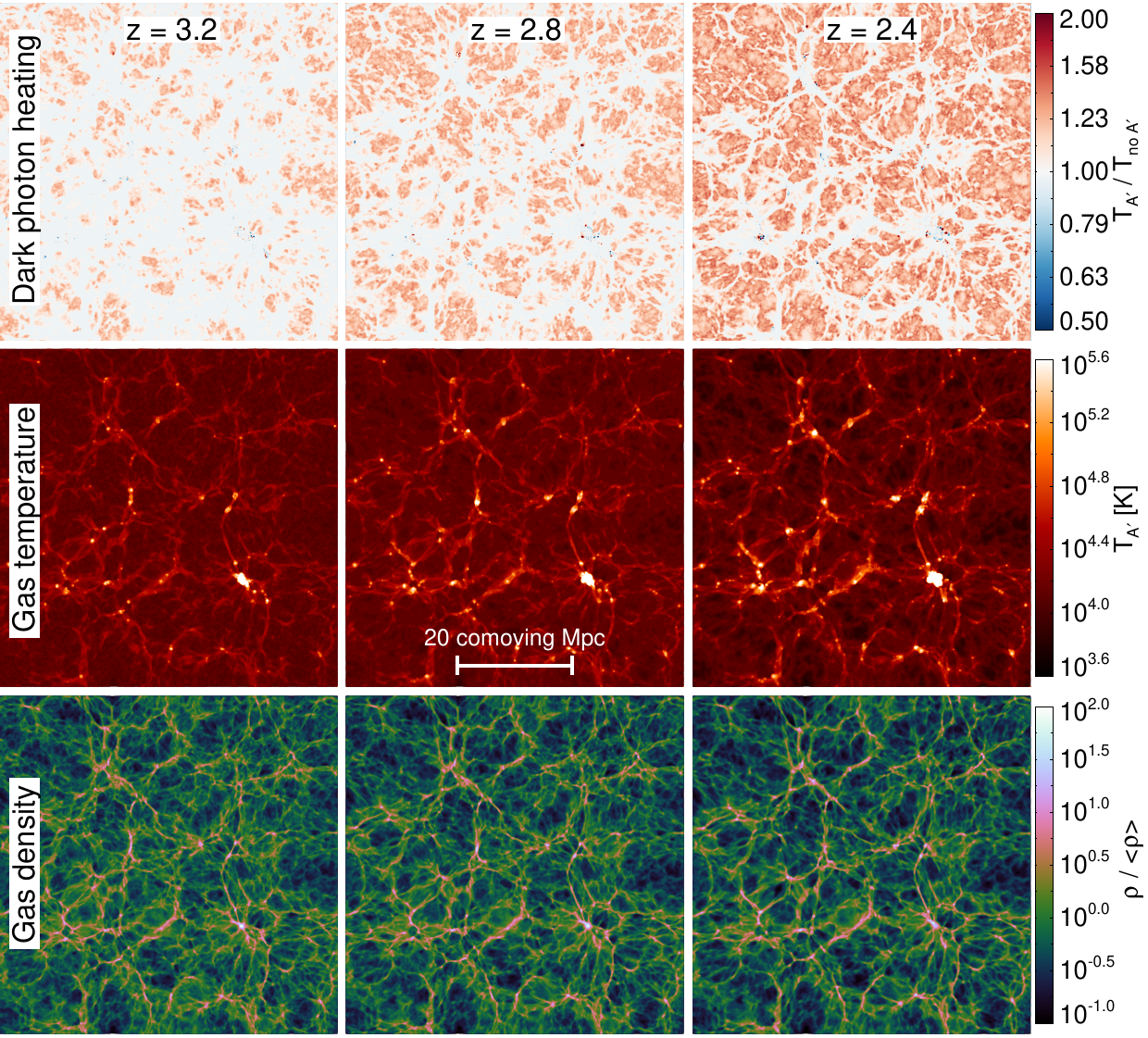}
    \caption{Slices ($500~h^{-1}\unit{kpc}$ thick) of a simulation box with \DP~heating of the best-fit model in Ref.~\cite{Bolton:2022hpt} ($m_{A'}=8\times10^{-14}~\unit{\eVc},\, \epsilon = 5\times10^{-15}$) at redshift $z=3.2$, $2.8$ and $2.4$ from left to right. Each panel is $40\, h^{-1}\unit{Mpc}$ on the side. The middle and bottom panels show the time evolution of the baryonic gas temperature and overdensity, respectively. The upper panels show the ratio of the gas temperature in the \DP~run relative to the scenario without extra heating. These panels highlight the amount of energy deposited by dark photon resonant conversions and the regions in the cosmic web that are affected by it.  Note the colour bar scales are all logarithmic.  Cosmic time evolves from left to right.}
    \label{fig:slabs}
\end{figure*}

Fig.~\ref{fig:slabs} shows slices of a simulated box that assumes the best-fit \DP~model proposed by Ref.~\cite{Bolton:2022hpt} ($m_{A'}=8\times10^{-14}~\unit{\eVc}, \, \epsilon=5\times10^{-15}$) at $z=3.2$, $2.8$ and $2.4$ from left to right. Resonant conversion occurs at $z \sim 1.75$ for mean density regions. 
The middle and bottom panels show the gas temperature and overdensity distributions. In contrast, the top panels show the ratio between the gas temperature with the \DP~model and the baseline \lcdm~scenario without exotic heating, highlighting the regions affected by dark photon resonant conversions and the amount of energy injected into the IGM. 
The plot shows that the \DP~model proposed by Ref.~\cite{Bolton:2022hpt}, in the considered redshift range, modifies the temperature of the baryonic gas up to a factor $\sim 2$ exclusively in the underdense regions where $\log \Delta\lesssim-0.5$ (e.g.\ voids) while not affecting the gas in overdense regions (e.g.\ filaments and nodes). 
Due to the redshift evolution of the free electron density $n_e$, given a specific mass, \DP~converts into photons in increasingly dense gas regions as cosmic time passes. Eventually, for the model shown, filaments and knots will be heated around $z\sim0$. 

\begin{figure*}[t!]
    \centering
\includegraphics[width=\linewidth]{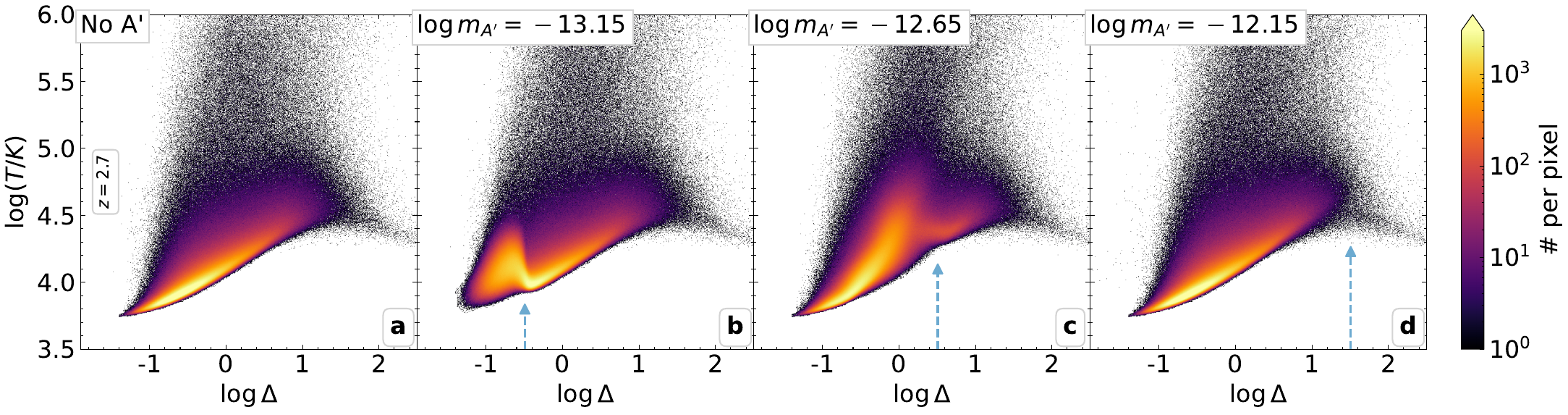}
    \caption{Distribution of the volume weighted gas temperature and overdensities $\Delta = \rho/\bar{\rho}$, in terms of mean density, of different models at redshift $z=2.7$. \textbf{a)} Baryonic gas distribution in the baseline \lcdm~model without exotic heating. The core of the distribution at low densities is well described by a power law of the form $T=T_0\Delta^{\gamma -1}$ with $\gamma=1.3$. \textbf{b - d)} Temperature-density distribution of gas in models with \DP~heating, where dark photons have masses of $\log (m_{A'}/\unit{\eVc})=-13.15$ (b), $-12.65$ (c) and $-12.15$ (d). All three models assume a kinetic mixing $\log \epsilon = -14.0$.
    \DP~heating induces a sharp increase in the gas temperature at the overdensities for which the resonant condition is met. This overdensity is highlighted with a dashed blue arrow, scaling as $\Delta_{\rm res}\propto m_{A'}^2(1+z)^{-3}$. The amplitude of the temperature increase depends on the amount of energy injected into the gas due to dark photon oscillations, scaling as $\Delta T_{\rm res} \propto E_{A'\rightarrow\gamma}\propto m_{A'}\epsilon^2(1+z)^{-3/2}$. For a fixed $m_{A'}$, a weaker kinetic mixing produces a less severe temperature feature, maintaining its density position.
    At the limit, models with $\log\epsilon \ll -14$ are not distinguishable from the model without extra heating, regardless of $m_{A'}$. 
    }
    \label{fig:Trho}
\end{figure*}

Fig.~\ref{fig:Trho} shows the distribution of the volume weighted temperature as a function of overdensity for the baseline \lcdm~model without extra heating and three \DP~scenarios at redshift $z=2.7$. The three models shown have different $A'$ masses but the same kinetic mixing $\log \epsilon = -14.0$. 

The presence of \DP~heating can be summarized as a sudden increase, relative to the baseline model, of the temperature distribution in a narrow density interval, corresponding to regions where the resonance condition was met at or just before the observed redshift ($m^2_\gamma(z,\textbf{x}) = m_{A'}^2$). The overdensity position of the resonance feature, $\Delta_{\rm res}$, depends solely on $A'$ mass and redshift, scaling as  $\Delta_{\rm res} \propto m_{A'}^2(1+z)^{-3}$.
Increasing the kinetic mixing $\epsilon$ increases the amplitude of the temperature feature, scaling as $\Delta T \propto E_{A'\rightarrow\gamma} \propto m_{A'}\epsilon ^2 (1+z)^{-3/2}$. Therefore, models with weaker mixing than those shown in Fig.~\ref{fig:Trho}, produce a smaller temperature increase.

The distribution shows that the IGM is strongly sensitive to the presence of $A'$ DM, displaying an anomalous and measurable thermal excess proportional to the amount of energy injected in the gas due to dark photon conversions, making it an appealing proxy to probe the presence of $A'$ DM. 

We notice that if the \DP~mass is such that a resonance never happens within the gas densities probed by the \lya forest, then we do not see any significant effect. Thus we expect to be sensitive only to a restricted range of \DP~ masses; higher or lower masses may be accessed with other probes~\cite{Caputo:2021eaa}.

\subsection{Discovering $A'$ DM with the Lyman-$\alpha$ forest transmitted flux}

The presence of $A'$ DM may impact the \lya forest in the spectra of distant bright quasars, caused by intergalactic neutral hydrogen gas. The shape of the absorption lines is directly related to the thermodynamic state of the absorbing gas, and the presence of \DP~heating will therefore modify the line widths and depths in a unique way, depending on the \DP~mass and kinetic mixing. 
In this section, we carry out a complete comparison between synthetic spectra extracted from the simulation boxes and a high-resolution, high-signal-to-noise observed spectrum of the \lya forest.  
We exploit the distribution of flux values across the sightlines, which is sensitive to the presence of \DP~heating, to compare data and simulations.

We focus on the redshift range $z>2$ for which the \lya forest is accessible by ground-based telescopes. For lower redshifts, the \lya transition falls below the atmospheric UV absorption cut-off ($\lambda<300$ nm). Moreover, we limit our analysis to $z<4$, as at higher redshifts the amount of neutral hydrogen in the IGM increases, leading to a tightly crowded and blended \lya forest in the spectra of high redshift quasars. In this instance, a proper assessment of absorption line profiles becomes complex, and the effects of \DP~heating become more difficult to detect.

\subsubsection{Simulated \lya spectra}
From the simulation boxes, we produce synthetic quasar sightlines by extracting 5000 pencil-beam skewers piercing the simulation volume, at redshift intervals of $\Delta z = 0.1$, each divided into 2048 bins along its length, saving the gas temperature, gas overdensity, local peculiar velocity and neutral hydrogen fraction.  The skewers are then used to compute the \lya optical depth, $\tau_{\rm Ly\alpha}$, as a function of wavelength, using the interpolation scheme of Ref.~\cite{Theuns1998} combined with the Voigt profile approximation from Ref.~\cite{TepperGarcia:2006mg}. 
% Finally, we rescale the optical depths of all pixels in the sightlines by a constant to explore the parameter space $\bar{F} = 0.71 - 0.82$, around the observed mean transmission of the \lya forest from Ref.~\cite{Becker13}. 
% From the simulations, we produce synthetic quasar sightlines by extracting the gas particle data (density, temperature, peculiar velocity and HI fraction) along skewers piercing the cosmological box and computing the \lya optical depth, $\tau_{\rm Ly\alpha}$, as a function of wavelength, following Refs.~\cite{Bolton:2022hpt, Bolton2022} (see the Methods section for more details). 
The normalized transmitted flux, i.e.\ the quasar flux fraction not absorbed by the neutral hydrogen, is defined as $F = e^{-\tau_{\rm Ly\alpha}}$. 

\begin{figure*}[t!]
    \centering
    \includegraphics[width=\textwidth]{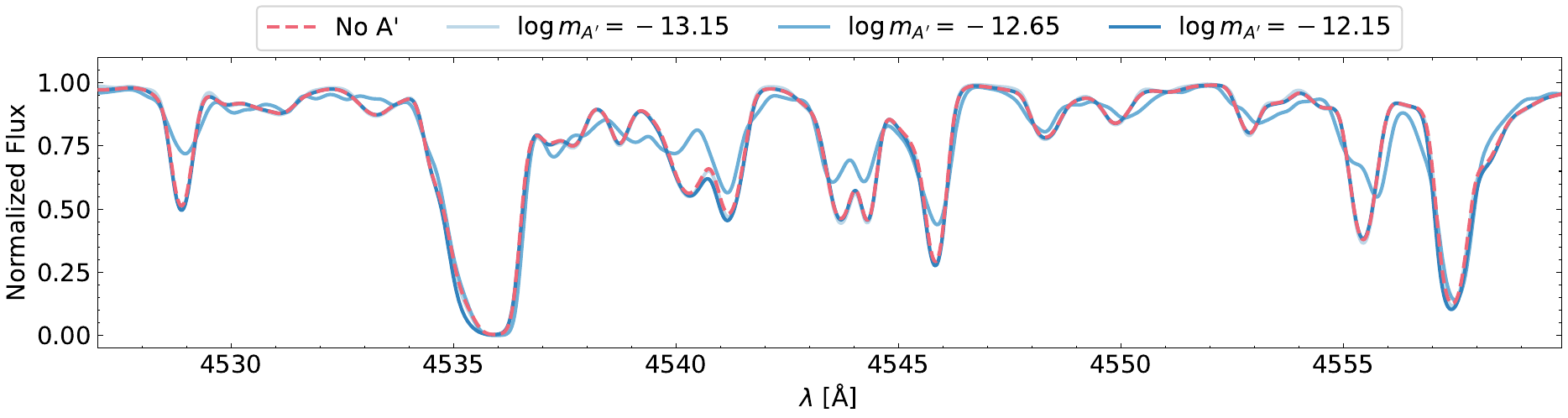}
    \caption{
    Section of a synthetic noiseless sightline extracted from different simulated boxes with and without \DP~heating. The dashed pink line is the baseline \lcdm~case without $A'$. Blue shaded lines show the transmitted fluxes with \DP~heating, where darker tones represent more massive models. The three \DP~models reported are the same as in Fig.~\ref{fig:Trho} and all assume strong kinetic mixing ($\log\epsilon=-14.0$). Only one model visibly differs significantly from the no-heating case ($\log m_{A'}/~\unit{\eVc}=-12.65 $) and is ruled out by our later analysis. 
    }
    \label{fig:flux_reg}
\end{figure*}

Fig.~\ref{fig:flux_reg} shows an example of the simulated normalized transmitted flux at $z=2.7$, extracted from the same models reported in Fig.~\ref{fig:Trho}. The baseline model without extra heating due to \DP~is shown as a dashed pink line, whereas the three models with \DP~are shown with blue shaded lines, with darker tones denoting heavier masses. 
The excess heating due to \DP~resonant conversions translates into the broadening of the \lya lines associated with gas particles having overdensities close to the resonance density $\Delta_{\rm res}$. 

For light mass \DP~ models (e.g. $\log(m_{A'}/~\unit{\eVc})=-13.15$), the resonance condition happens in underdense gas at $\Delta_{\rm res} \ll 1$ at the redshift in consideration, $z=2.7$. This low density gas absorbs only a small fraction of the incoming flux, and thus any modification of its thermodynamic state only affects the weak \lya lines close to the continuum, leading to a spectral shape that is not distinguishable by visual inspection from the baseline model without exotic heating. 

On the other hand, for intermediate masses (e.g. $\log(m_{A'}/~\unit{\eVc})=-12.65$) the conversion occurs at $\Delta_{\rm res}\sim1$. This gas is sufficiently dense to leave many strong, yet not saturated, lines in the spectrum, and an exotic energy injection leads to a visible difference with respect to the baseline no-$A'$ model. 

Lastly, more massive \DP~models (e.g. $\log(m_{A'}/~\unit{\eVc})=-12.15$) inject energy into gas at $\Delta_{\rm res} \gg 1$. This gas is mainly associated with very dense structures, such as filaments and knots in the cosmic web, whose absorption lines are deep, saturated and less likely to appear in a spectrum, due to the reduced number and relatively small cross-section of such structures. Moreover, the profile of saturated lines, i.e.\ lines that absorb 100\% of the incoming photons around the transition's wavelength, is not as sensitive to a change in gas temperature as for non-saturated lines, limiting the detectable effects of \DP~heating. The spectrum of such massive \DP~models only marginally differs from the no-$A'$ case. 

Note however, that the spectra shown in Fig.~\ref{fig:flux_reg} are noiseless and with infinite spectral resolution, i.e. are not affected by instrumental defects that will further reduce the differences between \DP~and no-$A'$ models. All three \DP~models shown assume a kinetic mixing $\log \epsilon=-14.0$. Models with weaker mixing show lines with a less severe anomalous broadening, down to the limit of producing the same normalized transmitted flux as the model without \DP~heating, regardless of $m_{A'}$.

\subsubsection{Data and mock spectra}
The presence of \DP~heating in the IGM, at the redshift in consideration,  broadens the weak and non-saturated \lya absorption lines close to the quasar emission continuum. As shown in Fig.~\ref{fig:flux_reg}, we are chasing differences in the line profile shape with and without $A'$ that may be fairly small. Therefore, in order to be able to detect (or constrain) the presence of exotic \DP~heating, we ought to analyse high-resolution quasar spectra with very high S/N levels. 

With this in mind, we analyse the high-resolution spectrum of the quasar HE0940-1050 ($z_\text{em}=3.0932$)~\cite{Rorai17} observed with VLT/UVES~\cite{Dekker2000} as the brightest object in the UVES Large Program~\cite{Bergeron2004}, having an average S/N in the \lya forest of 280 per resolution element. This is the high-resolution spectrum of a quasar at $2<z<3$ with the highest S/N found in the literature and thus is the best candidate for the following analysis. 

This spectrum has been used by Ref.~\cite{Rorai17} to constrain the thermal properties of the underdense IGM, through a novel manipulation of the normalized flux. 
By regulating the flux continuum level and rescaling the \lya optical depth before computing the distribution of the flux values of each pixel across the sightline, i.e. computing the regulated and transformed Probability Distribution Function (or tPDF throughout), Ref.~\cite{Rorai17} showed that it is possible to enhance the statistic's sensitivity to the thermal state of the underdense gas, stored in the high flux levels of the transmission spectrum. We carry out a similar analysis to constrain the presence of \DP~heating in the IGM, comparing the synthetic skewers extracted from the simulation outputs to the tPDF of HE0940-1050 \cite{Rorai17}.

Here we summarise the main steps of the procedure.
%%%%%%% mocks
First of all, we create a mock dataset of \lya sightlines that matches the redshift range of the Lyman forest found on the spectrum of HE0940-1050 (about $\SI{31000}{\km\per\second}$ long, equivalent to $\sim\SI{290}{Mpc}~h^{-1}$, centred at $z\sim2.75$). 
This long spectrum is composed by stitching multiple $\SI{40}{Mpc}~h^{-1}$ long sightlines extracted from simulation outputs at different redshifts, spanning the redshift range $z\sim 2.5 - 3$, to appropriately represent the redshift evolution of \DP~heating in the forest. 
We reproduce the instrumental effects on the synthetic sightlines by convolving the spectra with a Gaussian kernel with a full-width at half-maximum of $\SI{7.2}{\km\per\second}$ to match the UVES resolving power for $1$ arcsec slit aperture 
and then rebin the data onto a pixel grid with constant spacing $\Delta v=\SI{2.5}{\km\per\second}$. We reckon that the assumed spectral resolution only holds if the seeing at the moment of observation is smaller than the 1" slit, and could marginally change in the multiple exposures of the spectrum. In Sec~\ref{sect:ints_resolution} we investigate whether assuming a different spectral resolution for the mock spectra induces differences in the analysis results. 
% The resolution of the spectrum has also been estimated by fitting Gaussian profiles to the telluric absorption lines at $\sim6900$~\AA. These lines have a median FWHM of $\sim \SI{6.6}{\km\per\second}$, consistent with our assumed $\SI{7.2}{\km\per\second}$, accounting for the instrument's decrease in resolving power at the shorter wavelengths of the \lya forest that we are considering. 

The noise distribution of the original spectrum is modelled following the recipe of Ref.~\cite{Rorai17}. Summarizing, we 
\begin{enumerate}
    \item Divide the pixels of the original spectrum into 50 bins according to their flux values. The errors on each pixel produce 50 noise distributions for each flux value.
    \item To each simulated pixel, we assign a noise value randomly picked from the distribution associated with its flux level. 
\end{enumerate}

Lastly, since the mean transmitted flux value $\bar{F}$ of a \lya forest spectrum can exhibit large fluctuations among different sightlines, and can be rather sensitive to the number of strong absorption systems pierced by the sightline, it is reasonable to consider it as an independent free parameter within the measurement errors. This is particularly important when analysing a single sightline, as in our case, and not a large ensemble of spectra.
In this way we can put a constraint directly on the presence of \DP, marginalizing over cosmic variance effects. 
Therefore, we scale the mock flux skewers to a set of different mean flux values spanning the range $\bar{F} = 0.71 - 0.82$, around the observed mean transmitted flux at $z\sim2.7$ \cite{Becker13}.
We produce $N_{m}=10,000$ mock spectra for each $\{m_{A'},\epsilon,\bar{F}\}$ parameter point. 
The same procedure is carried out on the fiducial \lcdm~boxes without $A'$ heating.

%%%%%%%%%%%%%%%%% Flux regulation
Once we have a set of realistic data that matches the properties of the observed spectrum we move to the comparison between the two. 
We first note that the transmitted flux probability distribution function (PDF) is particularly sensitive to continuum placement uncertainty, whose erroneous estimation can lead to a shift of the flux levels and therefore to a stretch/compression of the probability distribution itself.

We reduce the uncertainty due to continuum placement by adopting a standard flux renormalization.
Taking the original fitted continuum as a starting point, the spectra are divided into regions of $\SI{10}{Mpc}~h^{-1}$. For each region, we find $F_{95}$, corresponding to the 95th percentile of the flux distribution within that region, and normalize the flux as $F_{r}=F/F_{95}$. This regulation aligns the spectra from simulations with different \DP~parameters at high fluxes, effectively removing any continuum bias placement in the PDF evaluation. 

%%%%%%%%%%%%%%% optical depth rescaling
Then, in order to focus on fluctuations near the continuum, we enhance the optical depth $\tau_{\rm Ly\alpha}$ of the \lya lines by an arbitrary constant factor $A$ as
\begin{equation}
    \tau_{\rm A} = A \,\tau_{\rm Ly\alpha},\,
\end{equation}
thus scaling the normalized (and regulated) flux as
\begin{equation}
    F_{\rm A} = \exp(A \ln\left| F_r \right|) =  \left| F_r \right|^A.
\end{equation}

We adopt a constant $A=10$ to compare our results to Ref.~\cite{Rorai17}.  This transformation is equivalent to changing the binning of the standard PDF in a flux-dependent way, such that the sensitivity at high fluxes is enhanced.
The optical depth transformation is applied to the continuum regulated flux $F_r$ described above. 

\begin{figure*}
    \centering
    \includegraphics[width=\linewidth]{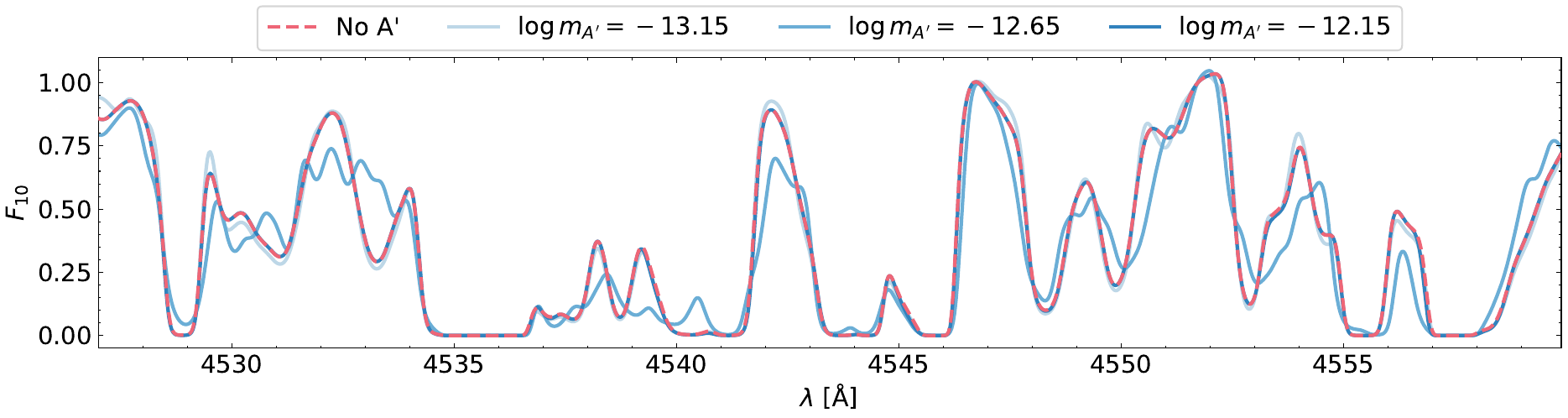}
    \caption{Same section of the synthetic sightlines showed in Fig.~\ref{fig:flux_reg}, after continuum regulation and optical depth rescaling have been applied. The dashed pink line is the baseline \lcdm~ case without $A'$. Blue shaded lines show the transmitted fluxes with $A'$ DM heating, where darker tones represent more massive models. The three $A'$ DM models reported are the same as in Figs.~\ref{fig:Trho} and \ref{fig:flux_reg}, and all assume strong kinetic mixing ($\log\epsilon = -14.0$). After the continuum regulation and optical depth transformation, the light $A'$ DM model ($\log( m_{A'}/~\unit{\eVc})=-13.15$) is visibly distinguishable from the no-$A'$ case.}
    \label{fig:flux_reg_rescaled}
\end{figure*}

Fig.~\ref{fig:flux_reg_rescaled} shows the regulated and rescaled flux, $F_{10}$, of the same noiseless sightline reported in Fig.~\ref{fig:flux_reg}. The procedure amplifies the differences among models and shows a visible distinction between the no-$A'$ and the light $A'$ DM model ($\log( m_{A'}/~\unit{\eVc})=-13.15$), which is negligible in the standard flux of Fig.~\ref{fig:flux_reg}.

Finally, we compute the probability distribution of $F_{10}$ throughout the mock datasets, called the regulated and transformed PDF, or tPDF for short. 

\begin{figure}[t!]
    \centering
    \includegraphics[width=\linewidth]{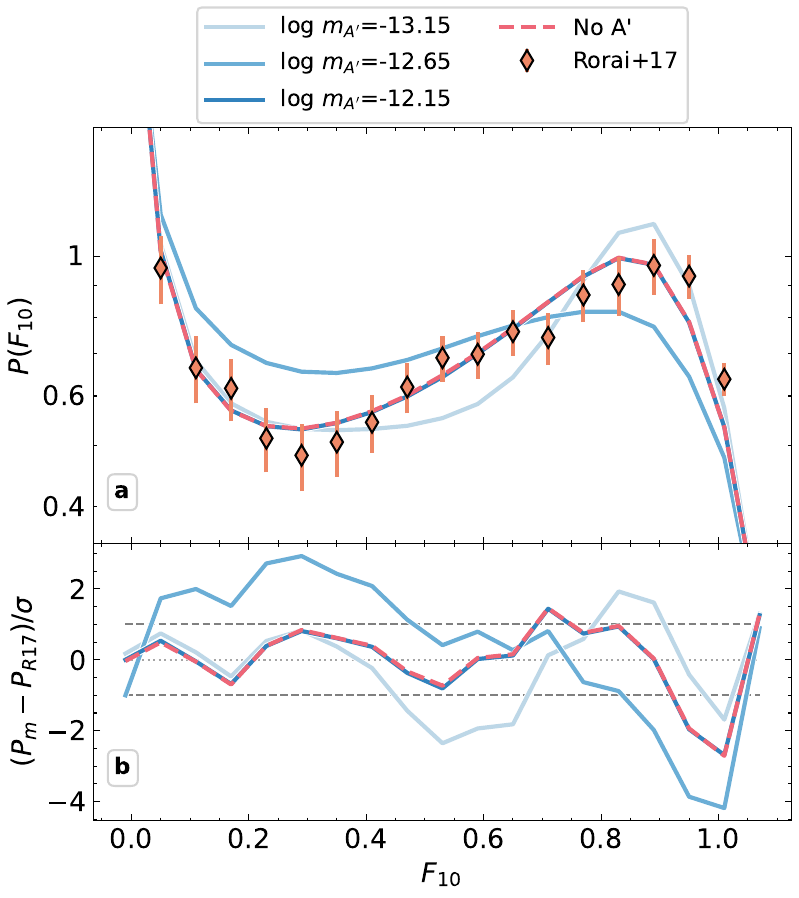}
    \caption{\textbf{a)} Regulated and transformed PDF (or tPDF) of mock spectra emulating the resolution, pixel size and noise distribution of the VLT/UVES spectrum of HE0940-1050, whose tPDF is reported in orange diamonds with error bars, as computed by \cite{Rorai17}. The tPDF of the same models reported in Figs.~\ref{fig:Trho} - \ref{fig:flux_reg_rescaled} are shown. All models reported have $\log\epsilon=-14.0$ and $\bar{F}=0.76$. \textbf{b)} The fractional difference between each model's tPDF and the observed data, in terms of the data error estimate. The horizontal grey dashed lines define the 1$\sigma$ region around the data points reported in Ref.~\cite{Rorai17}. }
    \label{fig:tPDF}
\end{figure}

Fig.~\ref{fig:tPDF} shows the tPDF of the same \DP~models reported in Figs.~\ref{fig:Trho} - \ref{fig:flux_reg_rescaled}, compared to the tPDF of HE0940-1050 \cite{Rorai17} shown as orange diamonds with error bars. The dashed pink line shows the tPDF of the baseline no-$A'$ model, whereas the \DP~models are shown in blue shaded solid lines, with darker tones denoting heavier \DP~masses.  
Most importantly, we see that the tPDF is able to distinguish between different \DP~prescriptions up to several $\sigma$, where a change in $m_{A'}$ modifies the shape of the tPDF. Changes in $\epsilon$ and $\bar{F}$ modify the tPDF shape in a non-degenerate way (not shown for conciseness). For low values of $\epsilon$, the energy injection due to \DP~oscillation is effectively negligible and the tPDF is indistinguishable from the no-$A'$ case, regardless of mass.
As expected, the massive case ($\log( m_{A'}/~\unit{\eVc})=-12.15$) shows the same tPDF as the baseline model, since it induces heating only in very dense and rare environments, affecting a relatively small number of lines and thus having small statistical impact on the full spectrum flux distribution.  
Notably, the IGM at $z \sim 2.7$ shows good agreement with our no-$A'$ simulation results.
% \HL{needs some discussion of the plot, explaining the result.}

\subsubsection{Likelihood}
% We compare the mocks tPDF to the data of \cite{Rorai17} employing a likelihood analysis. 
Following the standard assumption that the 19 flux bins in the tPDF are distributed as a multivariate Gaussian, we compute the likelihood between the model and data tPDF for each set of mock sightlines through the function
\begin{equation}\label{eq:like}
    \mathcal{L}(d|m_{A'},\epsilon; \bar{F}) = \frac{1}{\sqrt{\left( 2\pi \right)^k \left|\Sigma\right|}}
    \exp{\left[-\frac{1}{2} \Delta \textbf{P}^T \Sigma^{-1} \Delta \textbf{P} \right]} \,,
\end{equation}
where $\Delta \textbf{P} = \textbf{P}(m_A',\epsilon;\bar{F}) - \textbf{P}_d$ is the difference between the average tPDF of the mock spectra, extracted from the \DP~model with mass $m_{A'}$, kinetic mixing $\epsilon$ and scaled to mean flux $\bar{F}$, and the measured tPDF $\textbf{P}_d$ of Ref.~\cite{Rorai17}. $k$ is the number of tPDF bins considered, $\left|\Sigma\right|$  and $\Sigma^{-1}$ denote the determinant and the inverse of the covariance matrix, respectively, whose $ij$-th element is computed as %\HL{shouldn't the summation be $\sum_{n = 1}^{N_m}$?}
\begin{equation}\label{eq:covariance}
    \Sigma_{i,j}(m_{A'},\epsilon; \bar{F}) = \frac{1}{N_{m}}\sum_{n}^{N_{m}}\left(P_{i,n} - \bar{P}_i\right)\left(P_{j,n} - \bar{P}_j\right) \,,
\end{equation}
where $P_{i,n}$ and $\bar{P}_i$ are the tPDF values of the $i$th flux bin computed on the $n$th sightline and averaged on all mocks, respectively. The sum is performed over all $N_m$ mock spectra of the same model parameter combination.  We stress that the covariance matrix is model dependent, especially in the off-diagonal terms, and is recomputed on each \{$m_{A'},\epsilon,\bar{F}\}$ configuration.
Two degrees of freedom are removed by the tPDF normalization condition ($\sum P(F_i)=1$) and by the percentile continuum regulation performed on the flux skewers, hence we remove from our analysis the highest and lowest flux bins, following Ref.~\cite{Rorai17}.% \HL{can we say following Rorai+17 here?} 

\subsection{Posterior Analysis}

% We explore the posterior probability distribution of the \DP~parameters using a Monte Carlo Markov-Chain (MCMC) approach. We linearly interpolate the likelihood evaluated on the simulation grid points and obtain MCMC chains using the \texttt{emcee} python package \cite{ForemanMackey2013}.
% We consider two instances, assuming flat priors on mass, mixing and mean flux, 
% or flat priors on mass and kinetic mixing and a Gaussian prior on mean transmitted flux $\mathcal{N}_{\bar{F}}(0.7371, 0.01)$, based on the measurement of Ref.~\cite{Becker13}. We check for chain convergence using the Gelman-Rubin test~\cite{GelmanRubin1992}.
We explore the posterior probability distribution of the \DP~parameters using a Monte Carlo Markov-Chain (MCMC) approach. We linearly interpolate the likelihood evaluated on the simulation grid points and obtain MCMC chains using the \texttt{emcee} python package \cite{ForemanMackey2013}.
We consider two instances, assuming flat priors on $\log m_{A'}$, $\log\epsilon$ and $\bar{F}$;
or flat priors on $\log m_{A'}$ and $\log\epsilon$, with a Gaussian prior on mean transmitted flux $\mathcal{N}_{\bar{F}}(0.7371, 0.01)$, based on the measurement of Ref.~\cite{Becker13}. We check for chain convergence using the Gelman-Rubin test~\cite{GelmanRubin1992}.

\begin{figure}[t!]
    \centering
    \includegraphics[width=\linewidth]{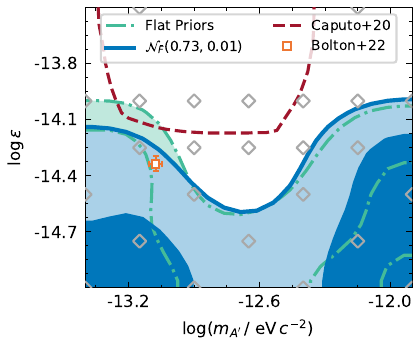}
    \caption{Posterior distribution over the \DP~mass - kinetic mixing plane, marginalized over the transmitted mean flux, assuming flat priors for each parameter (green dot-dashed line), and assuming flat priors on mass and mixing with a Gaussian prior on mean flux $\mathcal{N}_{\bar{F}}(0.7371, 0.01)$ \cite{Becker13} (blue area). The contours represent $1\sigma$ and $2\sigma$ levels. 
    Approximate constraints from Ref.~\cite{Caputo:2020rnx} are also shown (dashed red). The best fit \DP~model of Ref.~\cite{Bolton:2022hpt} is shown in orange, which is within $2\sigma$ of both posterior distributions. Grey squares denote the grid of simulated models used in this work.}
    \label{fig:posterior}
\end{figure}

Fig.~\ref{fig:posterior} shows the 1$\sigma$ and 2$\sigma$ levels of the posterior distribution in the mass-kinetic mixing plane, marginalized over transmitted mean flux, for the two choices of priors: all flat (dot-dashed green) and Gaussian over mean flux (solid blue). The contours in the two cases are similar, highlighting how the result is not prior dominated. The dashed red line reports the bounds on \DP~mass and mixing computed by \cite{Caputo:2020rnx} based on simple analytical energy distribution assumptions. The best fit model proposed by \cite{Bolton:2022hpt} is shown in orange and lies within the 2$\sigma$ region of both posteriors. 

The posterior distribution does not show a preference towards a particular \DP~model. Instead, the most likely models have either low kinetic mixing or very high mass. Both instances would produce a model that does not provide any non-negligible additional heating to the gas in the density range probed by the \lya forest, yielding a tPDF that is indistinguishable from the baseline no-$A'$ case within the error bars. Note moreover that the posterior distribution is limited, for high and low masses, at $\log\epsilon\sim-14.0$ due to the choice of parameter sampling we assumed and, requiring more simulations, would likely extend to higher mixings.

Overall, the analysed data does not demonstrate any substantial preference for \DP~heating.
% \AT{Discuss}

\subsection{Dark photon dark matter limits}
\begin{figure}[t!]
    \centering
    \includegraphics[width=\linewidth]{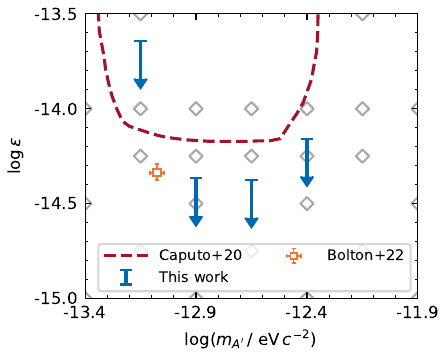}
    \caption{The 95\% confidence limits on the \DP~kinetic mixing parameter $\epsilon$ at fixed dark photon mass $m_{A'}$ arising from the analysis of the transformed flux tPDF at $z\sim 2.7$. The kinetic mixing at the high and low masses is not constrained by our data at $z\sim2.75$ within our parameter grid. 
    Approximate constraints from Ref.~\cite{Caputo:2020rnx} based on requiring that the total heat injected from \DP~not exceed \SI{1}{\eV} are also shown (dashed red). The best fit \DP~model of \cite{Bolton:2022hpt} is shown in orange. Grey squares denote the grid of simulated models used in this work.} 
    \label{fig:m-e_bounds}
\end{figure}

We next turn our attention to setting limits on $\epsilon$ as a function of $m_{A'}$. 
We consider the \lya forest mean flux as a nuisance parameter.
We linearly interpolate the likelihood values computed in Eq. \ref{eq:like} on a fine grid in the $\epsilon - \bar{F}$ plane, for a given fixed $m_{A'}$, and profile over the mean flux to obtain $\mathcal{L}(d|m_{A'},\epsilon)$, the maximum value of $\mathcal{L}(d|m_{A'},\epsilon;\bar{F})$ over all values of $\bar{F}$ for each pair of $(m_{A'}, \epsilon)$. We then define the test statistic
\begin{equation}
    \lambda(m_{A'}, \epsilon) = 2\left[\ln\mathcal{L}_{\max}(d|m_{A'}) - \ln\mathcal{L}(d|m_{A'},\epsilon)\right],
\end{equation}
where $\mathcal{L}_{\max}(d|m_{A'})$ is the maximum likelihood over all interpolated $\{\epsilon, \bar{F}\}$ points at a given mass $m_{A'}$. 
Wilks' theorem~\cite{Wilks:1938dza} then ensures that $\lambda$ follows a half-chi-square distribution~\cite{Cowan:2010js}.
We can therefore obtain the \DP~kinetic mixing 95\% confidence limits by finding the value of $\epsilon$ at which $\lambda = 2.71$. These limits are reported in Fig.~\ref{fig:m-e_bounds} for all mass values in our simulation ensemble, compared to the bounds obtained analytically by Ref.~\cite{Caputo:2020rnx}. 
Note that for low mass ($\log( m_{A'}/~\unit{\eVc})=-13.4$) and high mass models ($\log( m_{A'}/~\unit{\eVc})\ge-12.15$) we do not reach the 95\% confidence limit within the parameter space explored by our simulations. 

We also show the best-fit \DP~model of  Ref.~\cite{Bolton:2022hpt}, proposed to reconcile the discrepancy between the IGM temperature $z\sim0$ found in simulations and the one inferred by HST/COS observations. This model is allowed at 95\% C.L. by our bounds.

\section{Discussion}

In this work, we have carried out cosmological hydrodynamical simulations modelling the energy injection in the IGM due to dark photon dark matter resonantly converting into standard photons. From these simulations we have extracted synthetic \lya skewers and compared them to the ultra-high S/N VLT/UVES spectrum of quasar HE0940-1050 through the regulated and transformed flux PDF, setting robust constraints on the \DP~mass and kinetic mixing parameters.

For the first time, we utilize the full \lya forest in quasar spectra to constrain IGM heating from a DM process; almost all previous works do not account for inhomogeneities in the IGM and do not directly make use of \lya spectra, simply comparing a single, predicted IGM temperature with the gas temperature at mean density inferred from the \lya forest (see e.g.\ Refs.~\cite{Cirelli:2009bb,Diamanti:2013bia,Liu:2016cnk,Munoz:2017qpy,Liu:2020wqz}). 

We have obtained robust $95\%$ confidence limits on the kinetic mixing parameter of \DP~with mass around $m_{A'} \sim 10^{-13}-10^{-12} \, \rm eV/c^2$. This represents a significant improvement compared to previous work~\cite{Caputo:2020bdy,Witte:2020rvb}, which provided approximate limits. These limits were determined by requiring that the energy injected by dark photons during the HeII reionization epoch ($2\le z\le6$) did not cause an increase in the IGM temperature of more than  $\Delta T \lesssim 10^4 ~\unit{K}$. 

Our new results test the presence of \DP~with significantly more information than Ref.~\cite{Bolton:2022hpt}, since we are now comparing simulation results directly with \lya forest data at $z \sim 3$. We find that the best-fit point of Ref.~\cite{Bolton:2022hpt} is still consistent with the data; therefore, \DP~heating remains a viable solution to the discrepancy between HST/COS data and simulations at $z\sim0$, although we see no preference for the model either.

Low and high mass \DP~ models remain unconstrained by our data within the interval of kinetic mixings we consider, as these induce energy injection in a density regime not probed by the \lya forest. 
Nonetheless, these models can be investigated with the current data by expanding the parameter space to very strong mixing scenarios where large amounts of energy are injected into the IGM and can have a non-negligible effect on the \lya forest statistics, even if the oscillations occur in environments that are not well captured in the \lya forest. 
Similarly, \DP~models with heavy mass and strong mixing can alter the thermal state of the IGM at an earlier epoch and, if enough energy is injected, such trace will remain detectable at later times. In this work we do not extend the parameter space to high enough mixings to tackle this occurrence. 

Future deep spectroscopic observations will be required to explore the \DP~mass parameter space further, especially targeting the low redshift \lya forest down to $z\sim2$, where the best fit \DP~model of Ref.~\cite{Bolton:2022hpt} is expected to affect gas close to the mean density and so should be easier to detect and, eventually, confirm or rule-out. 
Such observations can already be carried out with current facilities and instrumentation (such as VLT/UVES \cite{Dekker2000}, VLT/ESPRESSO \cite{pepe2020ESPRESSO} and KECK/HIRES \cite{Vogt1994HIRES}), requiring, however, a substantial amount of integration time (of the order of a hundred hours), due to the low efficiency in the blue part of the spectrum of such instruments. 
In the future, spectrographs such as ANDES \cite{MARCONI22ANDES} will be more powerful and reach high S/N levels faster, being able to exploit the larger collecting area of the ELT, whereas, projects directly targeting the ultraviolet wavelengths such as VLT/CUBES \cite{ZanuttaCUBES} could push the search for \DP~down to $z\sim1.5$, probing even lighter models. 

We also note that dark photons in the mass range of interest for us can be produced around spinning black holes (BHs) through the superradiance instability, then affecting the spin-mass distribution of BHs or producing gravitational waves, for $m_{A'} \sim \SI{8e-14}{\eV}\,c^{-2}$~\cite{Baryakhtar:2017ngi,Cardoso:2018tly,Ghosh:2021zuf}. However, such constraints are currently subject to significant uncertainties~\cite{Belczynski:2021agb,Ghosh:2021zuf} and model dependence~\cite{Fukuda:2019ewf, Baryakhtar:2017ngi, Caputo:2021efm, Cannizzaro:2022xyw}. This highlights the importance of having complementary probes with distinct systematics in this mass range, such as the one presented in this work.

Finally, although we have focused here on one specific DM candidate, our pipeline can be applied to many other cases, such as decaying and annihilating dark matter, evaporating primordial black holes or axions converting in extra-galactic magnetic fields.  Our intention is to explore these other possibilities in the future.

\appendix
\section{Convergence of the transformed PDF}
\subsection{Instrumental resolution}\label{sect:ints_resolution}

The precise estimation of the instrumental resolution of the HE0940-1050 spectrum, especially in the \lya forest, can have a certain degree of uncertainty, as it is defined by the slit width and the sky seeing conditions at the moment of observations. As the spectrum used in our analysis is a composite of several archival observations, the final resolution is not straightforward to evaluate, and misestimation could impact the analysis outcomes.

Specifically, incorrectly assuming the width of the instrumental line spread function (LSF) alters the smoothing scale applied to the mock spectra, which might not reflect the real smoothing occurring in the observational data. This mismatch in smoothing and subsequent line broadening may be mistakenly attributed to a thermal effect from \DP, skewing the analysis results. 
To investigate this effect, we fitted Gaussian profiles to the telluric lines in the spectrum of HE0940-1050 at $\sim690$ nm. Since these lines are unresolved by the instrument, they serve to estimate the width of the spectrograph's line spread function and, by extension, its resolution. On average, the considered lines exhibit a FWHM of $\SI{6.6}{\km\per\second}$, which aligns closely with the fiducial FWHM of $\SI{7.2}{\km\per\second}$ used in our analysis during the mock creation process, accounting for the instrument's resolution degradation at the blue end of the spectrum. 

\begin{figure}[t!]
    \centering
    \includegraphics[width=\linewidth]{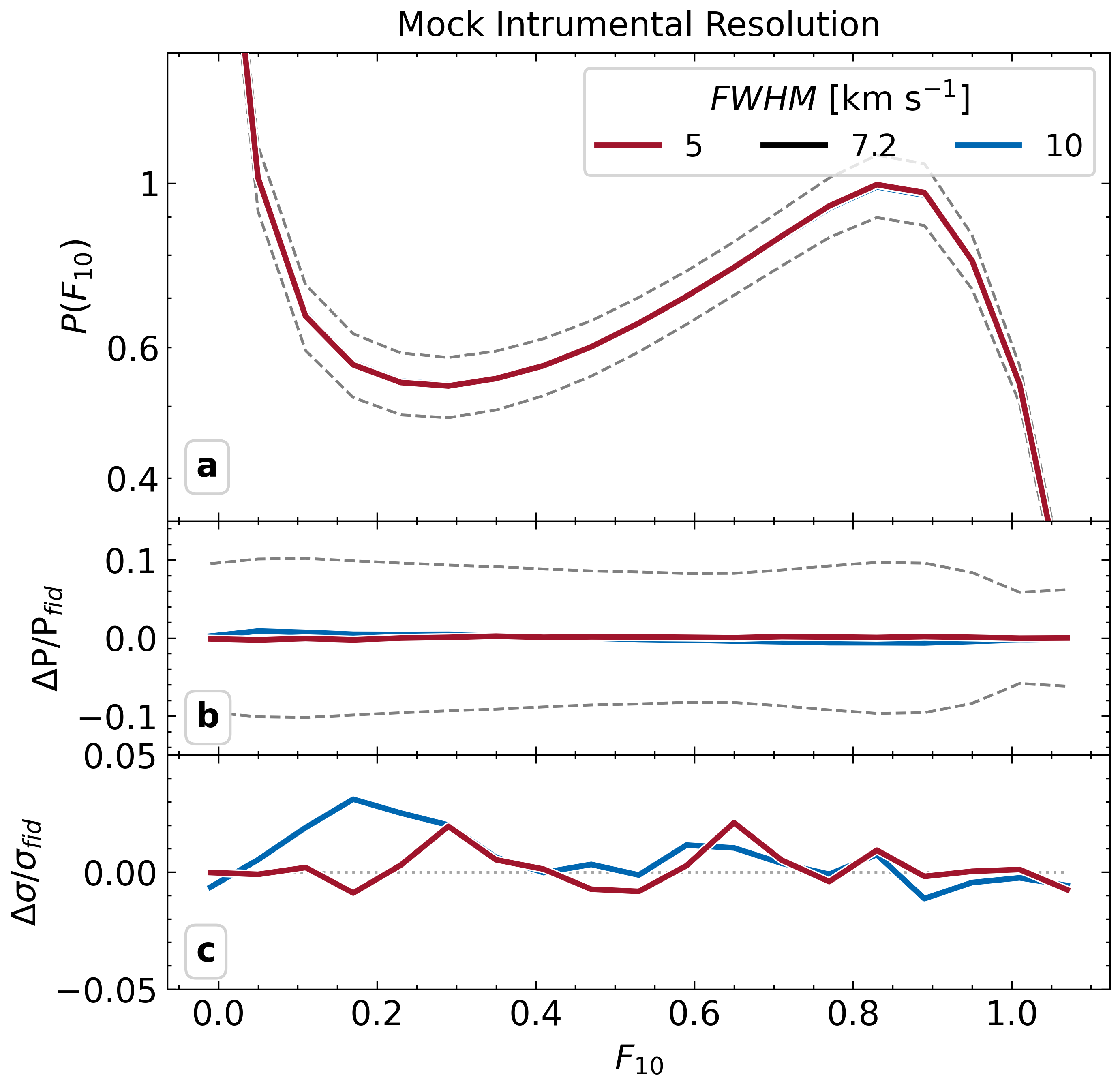}
    \caption{Convergence test of the tPDF with respect to the spectral resolution FWHM assumed in the mock spectra. Top panel: The tPDF of mock spectra, from the baseline no-$A'$ simulations, are shown, assuming a spectral resolution FHWM of $\SI{5}{\km\per\second}$ (red), $\SI{7.2}{\km\per\second}$ (black, fiducial) and $\SI{10}{\km\per\second}$ (blue). Dashed grey lines denote the 1$\sigma$ uncertainty on the fiducial tPDF. Middle panel: relative variation of the tPDF under different assumptions of spectral resolution, with respect to the fiducial $\SI{7.2}{\km\per\second}$ spectra. Bottom panel: relative variation of the tPDF error estimates, computed as the square root of the covariance matrix diagonal terms, with respect to the fiducial case.}
    \label{fig:FWHMres}
\end{figure}

Additionally, to effectively verify the absence of systematic errors in the \DP~ limit's estimation, we create a set of mock spectra derived from the simulations without exotic heating used in the main analysis, while assuming a FWHM of $\SI{5}{\km\per\second}$ and $\SI{10}{\km\per\second}$, and compute their tPDF. Fig~\ref{fig:FWHMres} displays the tPDF of the fiducial no-$A'$ mock dataset featuring a LSF with a FWHM of $\SI{7.2}{\km\per\second}$ (black), along with the two test sets with FWHM of $\SI{5}{\km\per\second}$ (red) and $\SI{10}{\km\per\second}$ (blue). The second panel illustrates the differences between the test mocks and the fiducial dataset, showing that the tPDF remains well-converged within uncertainties and is not impacted by the spectral resolution choice. Likewise, the tPDF uncertainty, depicted in the bottom panel as the square root of the diagonal terms of the covariance matrix, is converged within approximately $3\%$ when varying the LSF width. Therefore, we expect that the results of the analysis do not suffer from systematics under the choice of the fiducial spectral resolution assumed in the mock spectra, within realistic values of the FWHM, as estimated from the unresolved telluric lines. 

\subsection{Simulation mass resolution and box size convergence}\label{sec:resolution}

We assess the numerical convergence of the tPDF by varying the mass resolution and box size of the simulations. The fiducial no $A'$ model used throughout the work is kept as our reference point, with a box size $40~h^{-1}\unit{Mpc}$ and $2\times1024^3$ gas  and dark matter particles.  We contrast the tPDF predicted by this baseline run with those derived from two simulations featuring the same box size but with $2\times 768^3$ and $2\times 512^3$ particles, respectively.  This allows a check of convergence with simulation mass resolution.  Similarly, we check for convergence with simulation volume by comparing our fiducial run to a simulation with box size $20~h^{-1}\unit{Mpc}$ and $2\times 512^3$, particles, thus retaining the same mass resolution between the two runs. 

\begin{figure*}[t!]
    \centering
    \includegraphics[width=.9\linewidth]{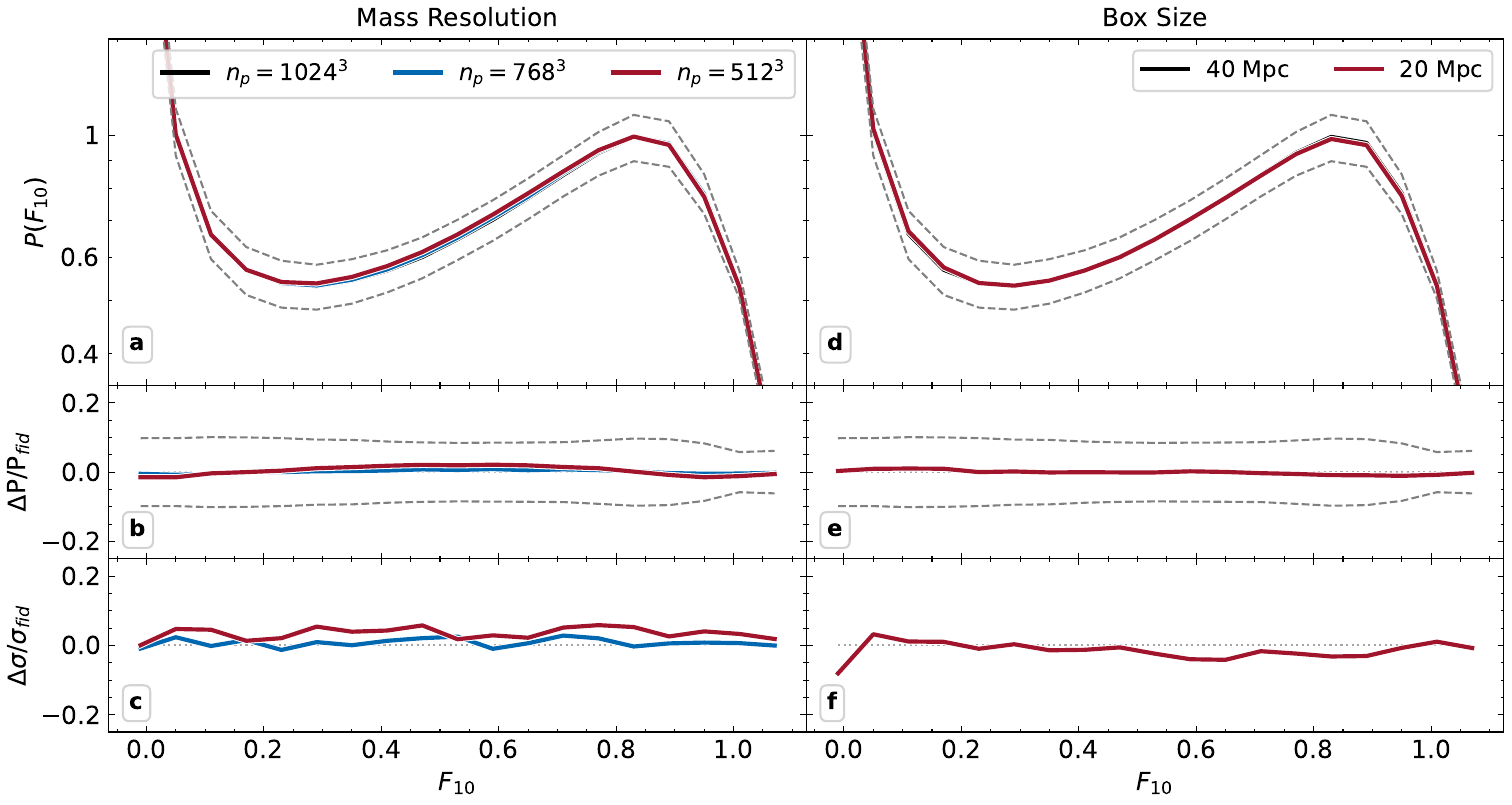}
     \caption{Convergence test of the tPDF with respect to mass resolution (left panels, a-c) and box size (right panels, d-f). The mass resolution test considers three simulations with the same box size (and initial conditions) and different numbers of particles: $2\times1024^3$ (our fiducial run), $2\times768^3$ and $2\times512^3$. The box size test considers two simulations with the same mass resolution but different box sizes: $40~h^{-1}\unit{Mpc}$ (fiducial run) and $20~h^{-1}\unit{Mpc}$. Top panels (a, d): The tPDF of the \lya forest in the no $A'$ scenario. The dashed grey line represents the error estimates on the fiducial simulation. Middle panels (b, e): relative variation of the tPDF with respect to the fiducial run. Bottom panels (d, f): relative variation in the tPDF error estimates, which are obtained from the square root of covariance matrix diagonal terms, with respect to the fiducial run.
    }
    \label{fig:resolconve}
\end{figure*}

Fig.~\ref{fig:resolconve} displays the tPDF of the simulations with different mass resolution (left panels) and box size (right panels).  The grey dashed curves are the uncertainties obtained from the square root of the diagonal elements of the simulated covariance matrix.   In all cases the tPDF is converged within the expected uncertainties.  The diagonal elements of the covariance matrix are also well converged, and are at most 6\% larger in the low-resolution runs. The covariance of the tPDF is only slightly influenced by box size, indicating that the baseline simulation adequately captures the cosmic web and does not suffer from large cosmic variance effects. In conclusion, we find our simulations are well converged.  We therefore do not consider any correction to the tPDF or its covariance to address poor convergence with either mass resolution or simulation volume.

\section*{Acknowledgements}
The authors thank Valentina D'Odorico for providing the original spectrum of HE0940-1050.  The hydrodynamical simulations used in this work were performed with the DiRAC Data Intensive service (CSD3) at the University of Cambridge, and the DiRAC Memory Intensive service Cosma8 at Durham University. CSD3 is managed by the University of Cambridge University Information Services on behalf of the STFC DiRAC HPC Facility (www.dirac.ac.uk). The DiRAC component of CSD3 at Cambridge was funded by BEIS, UKRI and STFC capital funding and STFC operations grants. Cosma8 is managed by the Institute for Computational Cosmology on behalf of the STFC DiRAC HPC Facility. The DiRAC service at Durham was funded by BEIS, UKRI and STFC capital funding, Durham University and STFC operations grants. DiRAC is part of the UKRI Digital Research Infrastructure.  JSB is supported by STFC consolidated grant ST/X000982/1. HL was supported by the Kavli Institute for Cosmological Physics and the University of Chicago through an endowment from the Kavli Foundation and its founder Fred Kavli, and Fermilab operated by the Fermi Research Alliance, LLC under contract DE-AC02-07CH11359 with the U.S. Department of Energy, Office of Science, Office of High-Energy Physics. The INAF authors acknowledge financial support of the Italian Ministry of University and Research with PRIN 201278X4FL, PRIN INAF 2019 "New Light on the Intergalactic Medium" and the ‘Progetti Premiali’ funding scheme.

\bibliographystyle{apsrev4-2}
\bibliography{DP_bib}

\end{document}